# Transmission of Voice Signal: BER Performance Analysis of Different FEC Schemes Based OFDM System over Various Channels

Md. Golam Rashed[1], M. Hasnat Kabir[2], Md. Selim Reza[3], Md. Matiqul Islam[2], Rifat Ara Shams[2], Saleh Masum[2], Sheikh Enayet Ullah[2]

[1]*Department of Electronics and Telecommunication Engineering (ETE)*
*Prime University, Dhaka-1216, Bangladesh.*
[2]*Department of Information and Communication Engineering,*
*University of Rajshahi, Rajshahi-6205, Bangladesh.*
[3]*Department of* Computer *Science and Engineering,*
*Atish Dipankar University of Science and Technology, Dhaka, Bangladesh.*
*golamrashed@primeuniversity.edu.bd*

***Abstract***

*In this paper, we investigate the impact of Forward Error Correction (FEC) codes namely Cyclic Redundancy Code and Convolution Code on the performance of OFDM wireless communication system for speech signal transmission over both AWGN and fading (Rayleigh and Rician) channels in term of Bit Error Probability. The simulation has been done in conjunction with QPSK digital modulation and compared with uncoded resultstal modulation. In the fading channels, it is found via computer simulation that the performance of the Convolution interleaved based OFDM systems outperform than that of CRC interleaved OFDM system as well as uncoded OFDM channels.*

## 1. Introduction

In last decades, Orthogonal Frequency Division Multiplexing (OFDM) based communication systems has been identified as one of key transmission techniques for next generation wireless communication systems [1]. The main attractions of OFDM are handling the multi-path interference, and mitigate inter-symbol interference (ISI) causing bit error rates in frequency selective fading environments [2]. Wireless mobile communication systems of the 21st century have to confirm a wide range of multimedia services such as speech, image, and data transmission with different and variable bit rates up to 2 Mbit/s [3]. It is all recognized that there is a great impact of channel coding on the performances of OFDM based wireless communication system to provide high data rates over severe multipath channels.

The ISI in OFDM based communication system can be eliminated by adding a guard interval which significantly simplifies the receiver structure. However, in order to take advantage of the diversity provided by the multi-path fading, appropriate frequency interleaving and channel coding is essential. Therefore, channel coding becomes an indivisible part in most OFDM system and a significant amount of research work has focused on optimum encoder, decoder, and interleaver design for information transmission via OFDM over fading environments [4-5]. On the other hand, some subcarriers of OFDM system may be completely lost because of deep fades. In this case, the overall performance will be largely dominated by a few subcarriers with small amplitudes. Error correction code can be used to avoid this domination by the weakest subcarriers. In OFDM system, several error-correcting codes have been applied such as convolutional codes, Reed-Solomon codes, Turbo codes [6], and so on. Forward-error correction (FEC) is one of the best for its potentiality.

The Signal-to-Noise Ratio (SNR) has been largely recognized as a performance analysis metrics over the past decade due to its simple mathematics [7–8]. Nevertheless, some other important performance evaluation metrics also have been used such as Bit Error Ratio (BER) [9] and Symbol Error Rate (SER) [10] which characterizes the associated performance degradation more accurately of the OFDM based communication systems.

Digital Audio Broadcasting (DAB) was the first commercial use of OFDM technology. DAB is a replacement for FM audio broadcasting, by providing high quality digital audio and information services. OFDM was used for DAB due to its multipath tolerance. Broadcast systems operate with potentially very long transmission distances (20-100 km). As a result, multipath is a major problem as it causes extensive ghosting of the transmission. This ghosting causes Inter-Symbol Interference (ISI), blurring the time domain signal. OFDM is a suitable candidate for high data rate transmission with forward error correction (FEC) methods over wireless channels [3].

In this paper, performance of different block codes namely CRC and Convolution, with Interleaver are analyzed using Differential QPSK modulation under the OFDM based wireless communication system over different channels (AWGN, Rician, and Rayleigh) on voice signal transmission. The object of this paper is to identify the behavior of interleaved convolutional code on OFDM based communication system and compare the results with other coded as well as uncoded results.

## 2. Related Works

K. Vivek et. al. has presented the bit error rate (BER) analysis of the coded OFDM communication systems [11]. Their simulation result showed that space time turbo coding with OFDM system is seen to provide maximum coding gain. On the other hand, In [12], authors consider the performance of coded OFDM using turbo-codes, for application in digital broadcasting systems. Authors showed that turbo codes can give performance improvements of some order of dB on a Rayleigh fading channel, over the conventional convolutional codes in the existing standards. Zahid Hasan concluded in [13] that the performance of the OFDM system in digital color image transmission over AWGN channel is comparatively better as compared with Rayleigh and Rician fading channels. Author also observed from his simulation that the performance of QPSK modulation technique much more better than BPSK because QPSK is double bandwidth efficiency thereby showing unique performance in proper identification and retrieval of transmitted digital image. M. K. Gupta, et.al. illustrates in [14] the way to increase the system throughput while maintaining system performance under desired bit error rate. From their study by simulation it is concluded that it is possible to improve the performance of uncoded OFDM can be improved by convolution coding scheme. In [15], the authors concluded that concatenated channel coding scheme with low-density parity-check and convolutional codes in OFDM based system is very much effective in proper identification and retrieval of transmitted digital image in noisy and fading environment. In [16], performance of Interleaved CRC encoded QPSK based wireless communication system are analyzed. Author found by simulation that interleaved CRC encoded QPSK based system provides unique performance in proper identification and retrieval of transmitted color image. Author further concluded that it is possible to transmit the color image with lower value of the transmitted power. Form the above results it is found that the performance of OFDM based communication system can be increased using suitable error correction code. including text, illustrations, and charts, must be kept within the parameters of the 8 15/16-inch (53.75 picas) column length and 5 15/16-inch (36 picas) column width. Please do not write or print outside of the column parameters. Margins are 1 5/16 of an inch on the sides (8 picas), 7/8 of an inch on the top (5.5 picas), and 1 3/16 of an inch on the bottom (7 picas).

## 3. FEC Codes and Interleaver

In telecommunication, forward error correction code (also called Error control coding) is widely used as an error controlling tool for data transmission. FEC incorporates with receiver. It has the ability to detect and correct a limited number of errors without needing a reverse channel to request retransmission of data. The application of FEC is suitable where retransmissions are relatively costly or impossible. There are several error controlling codes are available. However, popularly used Error control coding are Cyclic Redundancy Check (CRC) code and convolutional code. In this section first we discussed CRC and convolutional coding and then block Interleaving.

### 3.1. Cyclic Redundancy Check Coding

Cyclic Redundancy Code (CRC) provides the defense against data corruption in many digital networks. It detects accidental changes to raw computer data. Unfortunately, many commonly used CRC polynomials provide significantly less error detection capability than they might [17]. Mathematically, a CRC can be described as treating a binary data word as a polynomial over GF(2). A polynomial in GF(2) is a polynomial in a single variable x whose coefficients are 0 or 1. Addition and subtraction are done modulo 2—that is, they are both the same as the exclusive and operator. For example, the sum of the polynomials

$x^3 + x + 1$ and

$x^4 + x^3 + x^2 + x$ is $x^4 + x^2 + 1$, as is the their difference.

In the CRC method, a short and fixed-length certain binary number of check bits, often called checksum, are appended to the message beings transmitted. The receiver can determine whether or not the check bits agree with the data, to ascertain with a certain degree of probability whether or not an error occurred in transmission. An r-bit CRC checksum detects all burst errors of lengths ≤ r. A burst error of length r is a string of r bits in which the first and the last bit are in error, and the intermediate r-2 bits may or may not be in error.

CRC becomes a popular error correction coding technique due to several reasons such as simple to implement in binary hardware, easy to analyze mathematically, and particularly good at detecting common errors caused by noise in transmission channels. CRC is specifically designed to protect against common types of errors on communication channels, where they can provide quick and reasonable assurance of the integrity of messages delivered. However, they are not suitable for protecting against intentional alteration of data. However, as there is no authentication, an attacker can edit a message and recalculate the CRC without the substitution being detected.

### 3.2 Convolutional Coding

A Convolution encoder consists of a shift register which provides temporary storage and a shifting operation for the input bits and exclusive-OR logic circuits which generate the coded output from the bits currently held in the shift register. In general, k data bits may be shifted into the register at once, and n code bits generated. In practice, it is often the case that $k$=1 and $n$=2, giving rise to a rate ½ code. A rate ½ encoder illustrated in Figure 1[18] and this will be used to explain the encoding operation.

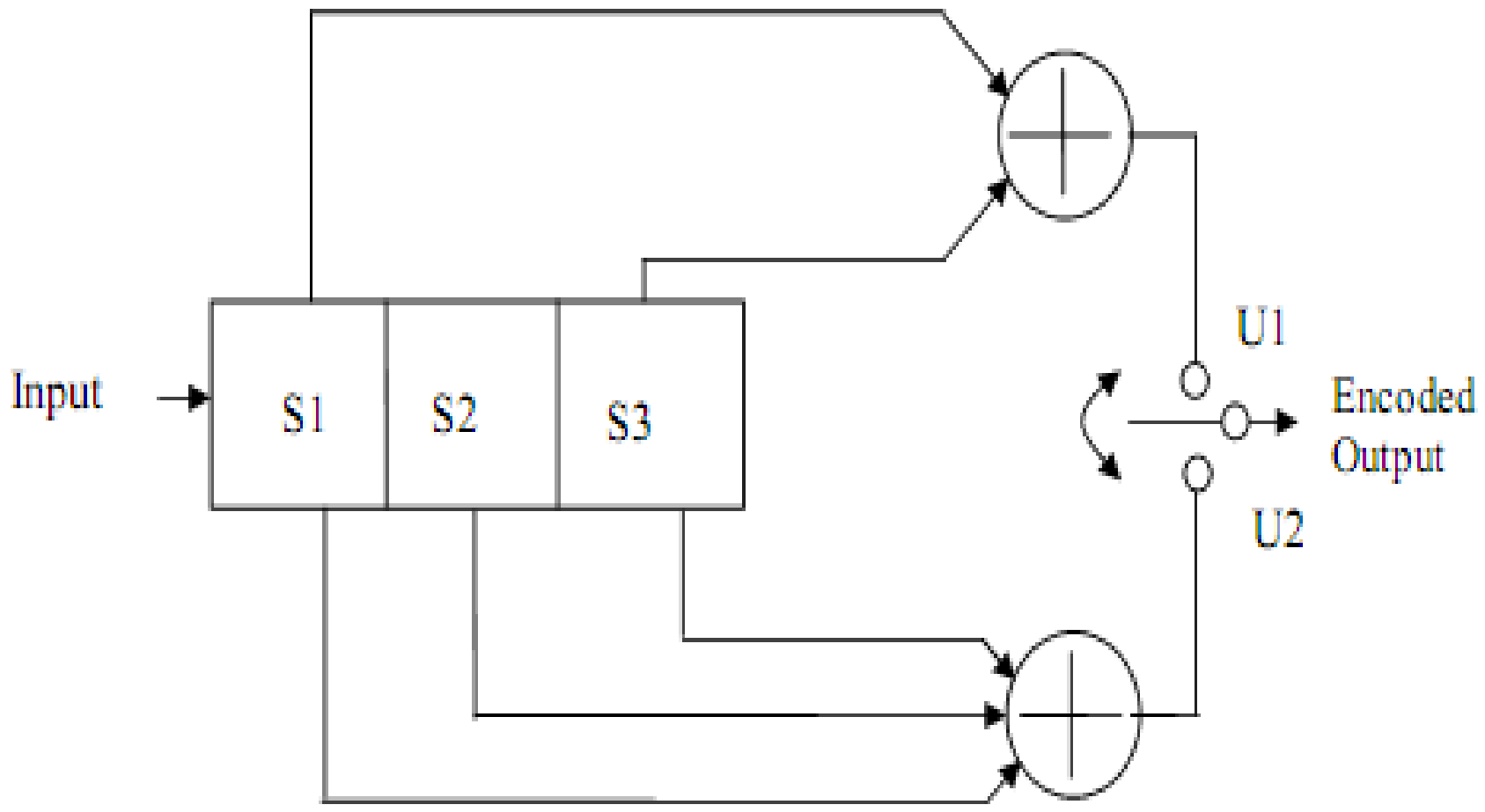

**Figure 1. A ½ rated Convolution Encoder**

Initially, the shift register holds all binary 0s. The input data bits are fed in continuously at a bit rate $R_b$, and the shift register is clocked at this rate. As the input moves through the register, the rightmost bit is shifted out so that there are always 3 bits held in the register. At the end of the message, three binary 0s are attached, to return the shift to its initial condition. The commutator at the output is switched at twice the input bit rate so that two output bits are generated for each input bit shifted in. At any one time the register holds 3 bits which from the input to the exclusive-OR circuits.

Convolutional codes have been widely applied to satellite communications. In cellular mobile communication, the channel characteristics is less favorable with burst errors arising from multipath(reflection), shadowing of the signal and co-channel interference, but the need to achieve coding gain at the moderate target bit error rates again dictates that convolutional codes should be used. Because of the hostile channel environment, the voice coders (vocoders) are designed to work well with bit error rate of $10^{-3}$ and acceptable with error rates well above this. Convolutional codes are highly suitable for AWGN channels [17]. In the case of GSM standard for digital mobile communication, convolutional codes are needed with interleaving to protect against the channel error burst.

### 3.3 Interleaving

In computer science and telecommunication, interleaving is a way to arrange data in a non-contiguous way to increase performance. It is typically used in error-correction coding, particularly within data transmission. Interleaving has become an extremely useful to transform analog voices into efficient digital messages that are transmitted over wireless links. Many communication channels are not memory less: errors typically occur in bursts rather than independently [19]. In multi-carrier communication systems, additional interleaving across carriers may be employed to mitigate the effects of prohibitive noise on a single or few specific carriers (e.g., frequency-selective fading in OFDM transmission) [20].

The function of the interleaver is to spread the source bits out in time so that if there is a deep fade or noise burst, the important bits from a block of source data are not corrupted at the same time. By spreading the source bits over time, it becomes possible to make use of error control coding (called channel coding) which protects the source data from corruption by the channel. Since error control codes are designed to protect against channel errors that may occur randomly or in a bursty manner, interleavers scramble the time order of source bits before they are channel coded. Examples of Inerleaving are stated below:

**Transmission without Interleaving:**

| | |
|---|---|
| Error-free message: | bbbbccccddddeeeeffffgggg |
| Transmission with a burst error: | bbbbcc_____deeeeffffgggg |

In the above case, the code-word ccddd is altered in four bits, so either it cannot be decoded at all or it might be decoded incorrectly.

**Transmission with Interleaving:**

| | |
|---|---|
| Error-free code-word: | bbbbccccddddeeeeffffgggg |
| Interleaved: | bcdefgbcdefgbcdefgbcdefg |
| Transmission with a burst error: | bcdefgbcdefgb____gbcdefg |
| Received code words after deinterleaving: | bbb_ccc_ddd_eee_fff_gggg |

In this case, the codewords aaaa, eeee, ffff, gggg, only one bit is altered, so one-bit error-correcting-code will decode everything correctly.

The latency is increased by interleaving techniques due to the entire interleaved block must be received before the packets can be decoded. Another advantage is provided by interleaver that it hides the structure of errors. Without an interleaver, more advanced decoding algorithms can take advantage of the error structure and achieve more reliable communication than a simpler decoder combined with an interleaver.

## 4. Simulation Model

The main objective of this work is to simulate the OFDM system by utilizing different interleaved FEC codes. The block diagram of the entire system is shown in Figure 2. The designated block represents their specific operation. We do not explain each block in details. Here we only consider two FEC codes namely Cyclic Redundancy Check (CRC) and Convolutional code (CC).

In order to compare the results through the simulation, we have considered a recorded voice signal as input signal to be transmitted over the OFDM based communication system. The numbers of bits are taken 64000. Figure 3 shows the graphical representation of a recorded audio signal of data length 8000 samples in 1 second. It is PCM encoded at a data rate of 64Kbps with sampling frequency 8000Hz and 8 bit A/D conversion.

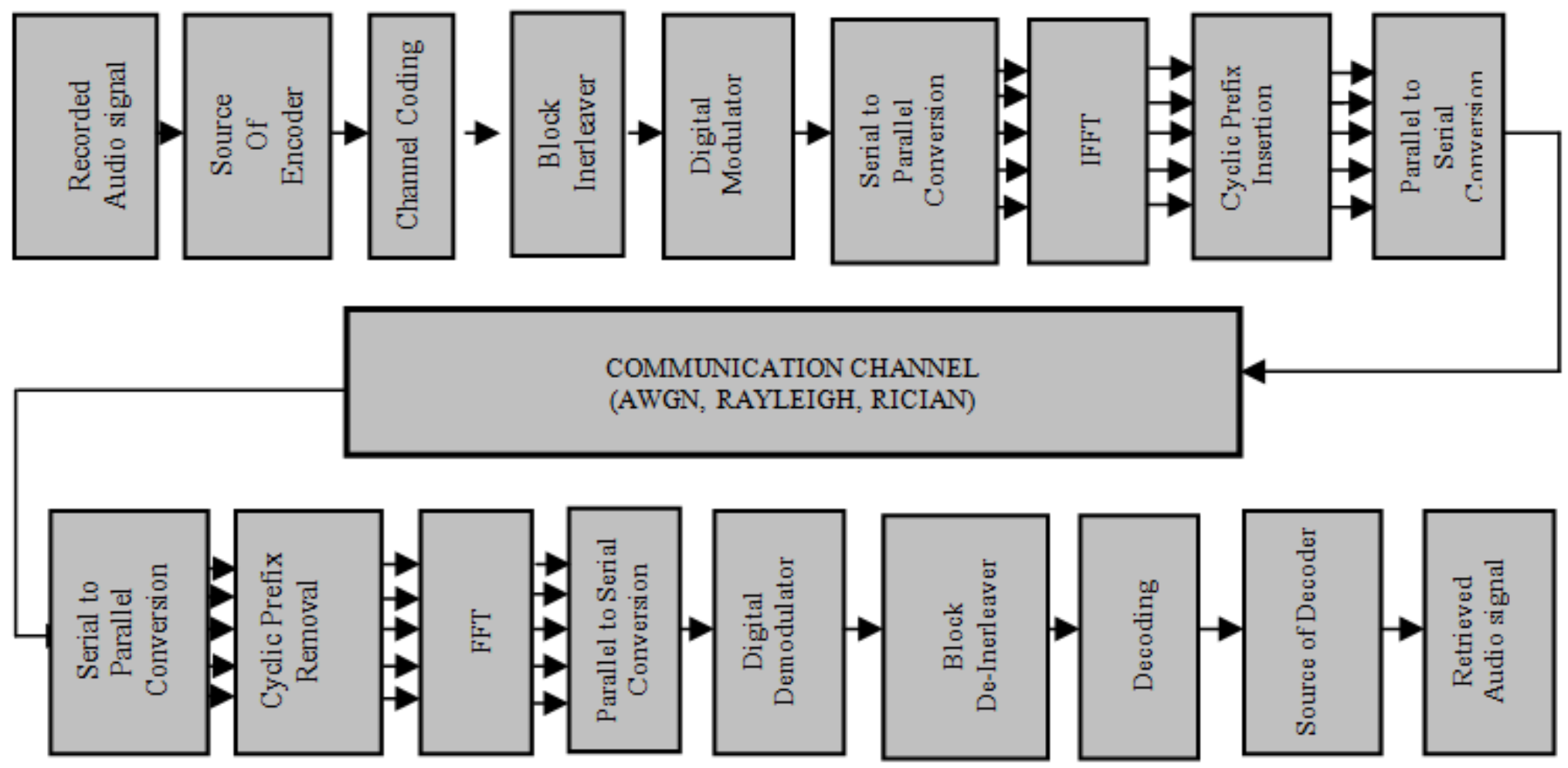

**Figure 2. OFDM Simulation model for transmission of recorded voice signal.**

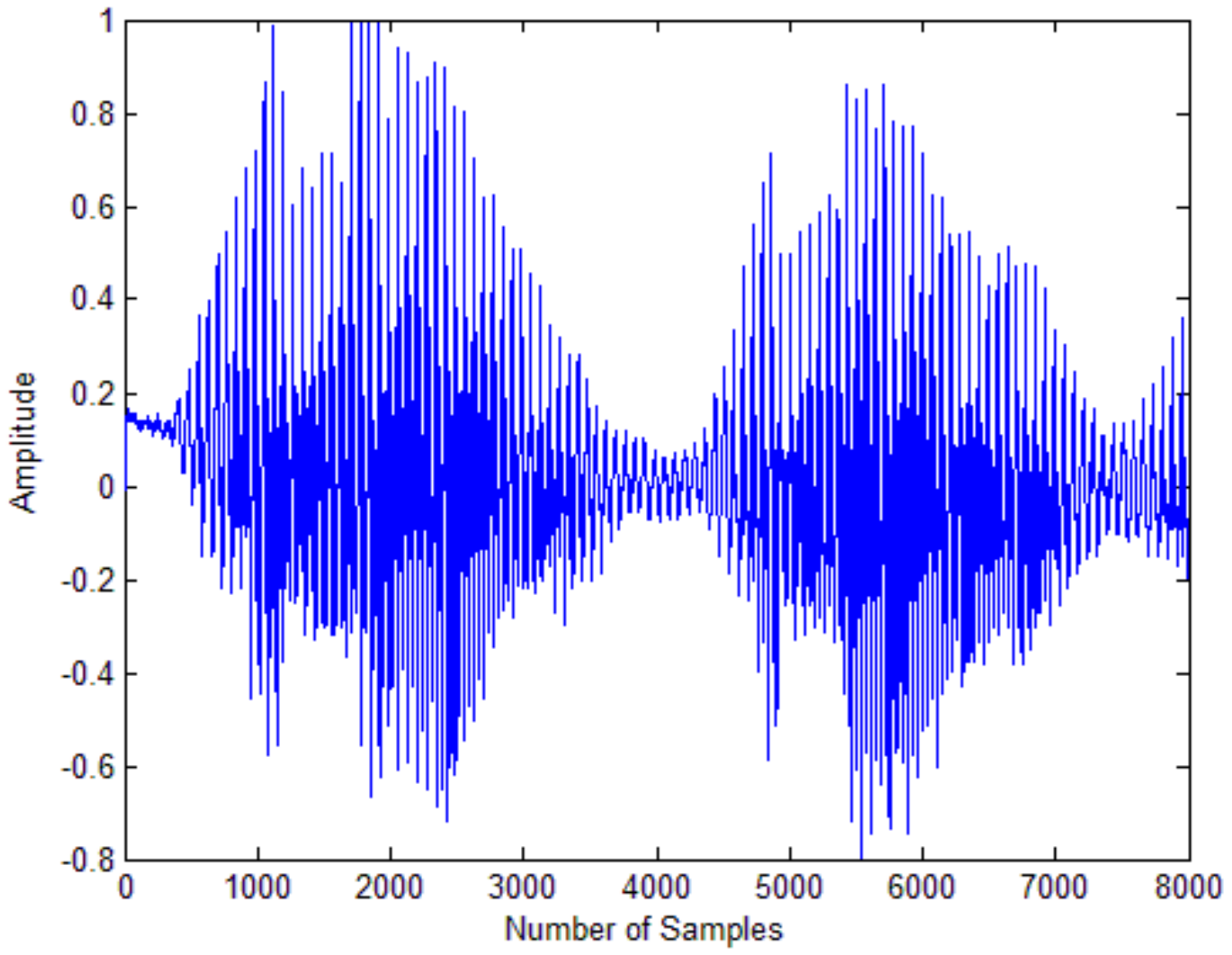

**Figure 3. Graphical Representation of Audio Signal**

MATLAB is used to write a computer program designed for simulation of an OFDM system to allow various parameters of the system to be varied and tested. The possible considered parameters are mentioned in Table 1 with their corresponding varied values.

**Table 1. Simulation Parameters.**

| Parameters | Values |
|---|---|
| Digital Modulation | QPSK |
| FEC codes | CRC, Convolutional |
| Convolutional code rate | ½ |
| Interleaver Size | 35×1 |
| SNR Range | 0 to 14 |
| Considered Channels | AWGN, Rayleigh, Rician |
| OFDM Block Size | 8 |
| Cyclic Prefix Length | 1 |
| Data Length | 8000 samples per second |
| Data Rate | 64 Kbps |
| Sampling Frequency | 8000 KHz |
| Code word Length | 8(for CRC) |
| Message Length | 2 (for CRC) |
| Constraint length of convolutional code | 7 |

A short description of system algorithm is listed below:

1. Consider a recorded audio segment as input signal.
2. Source encoding is used to transform the recorded audio signal into its corresponding information bits in order to represent the information in digital form.
3. Encode the information bits using either CRC or convolutional encoder with specified generator matrix.
4. Encoded information bits are then interleaved to obtain time diversity in a digital communications system without adding any overhead.
5. Use QPSK modulation to convert the binary bits, 0 and 1, into complex signals.
6. Performed serial to parallel conversion.
7. Use IFFT to generate OFDM signals.
8. Performed cyclic prefix insertion.
9. Introduce noise to simulate channel errors. We assume that the signals are transmitted over an AWGN channel and fading environment namely Rayleigh and Rician channels.
10. At the receiver side, perform reverse operations to decode the received sequences of information bits to retrieve the transmitted voice signal segment.
11. Count the number of specious bits by comparing the decoded bit sequences with original one.
12. Calculate the Bit Error Probability (BER) as a function of against different values of signal to noise ratio (SNR) and plot it accordingly.

## 5. Simulation Results and Discussion

In our study we have done all the simulations to achieve a desired Bit Error Probability. For analysis, we considered the AWGN channel, Rayleigh and Rician fading channel models. Parameters defined in section 4 are used to utilize above mentioned channels. The bit error probability performance of coded OFDM system is compared with the respective uncoded system under the AWGN and Fading (Rayleigh, Rician) channels. Figure 4 shows the performance of uncoded OFDM system in different channels in terms of bit error probability.

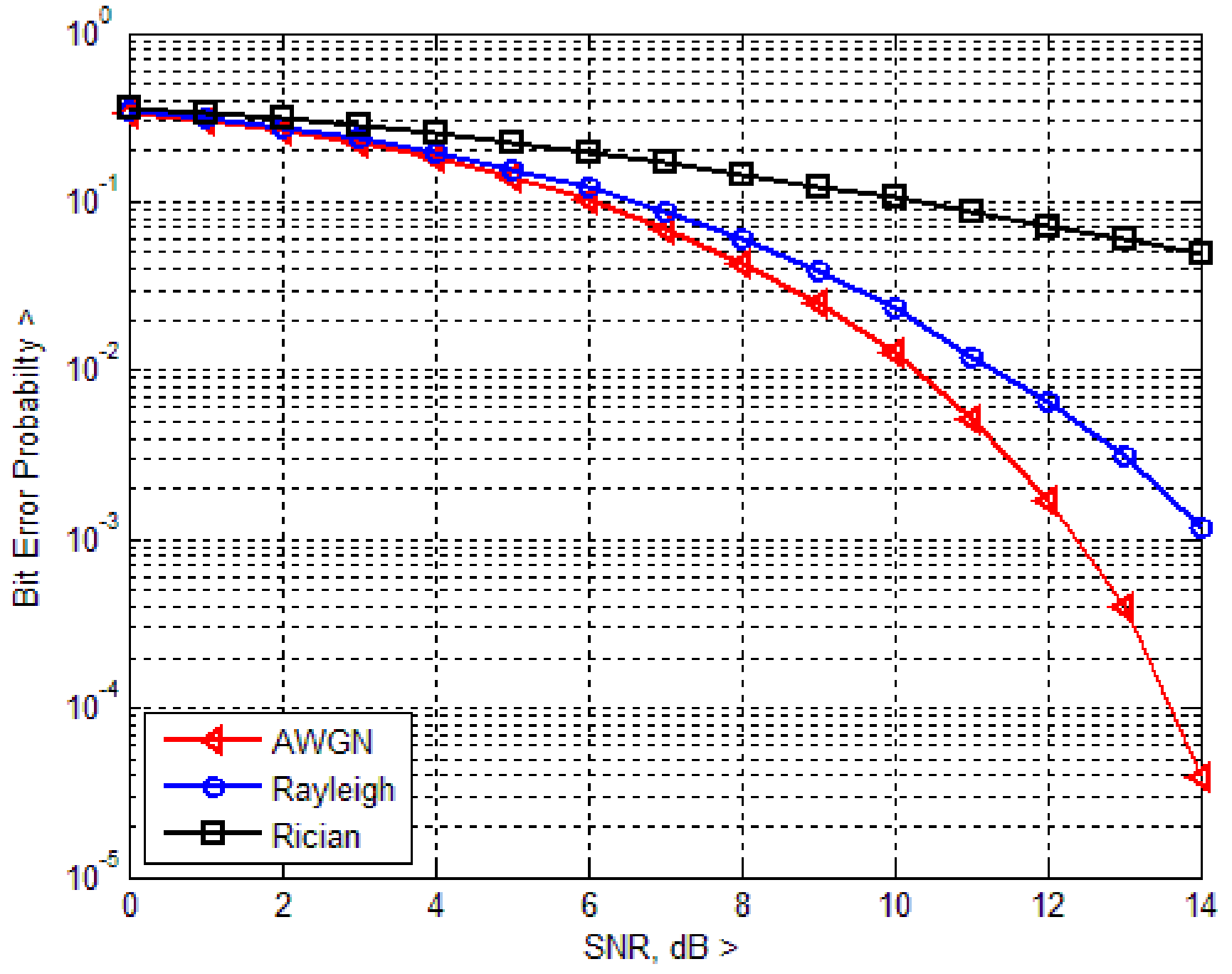

**Figure 4. Performance of uncoded OFDM system in different channels.**

In this figure, it is seen that, the bit error probability of AWGN channel gives better performance as compared with Rayleigh as well as Rician channel. The BER is inversely proportional to the signal to noise ratio (SNR). However, AWGN gives a desirable gain for higher value of SNR about 8 dB whereas gain is not so significant at lower value of SNR between 0 to < 8 dB. On the other hand, in comparison between Rayleigh and Rician fading channel, the system performance is better in Rayleigh fading channel than that of Rician. For a typical value of signal to noise ratio (SNR) 10 dB, approximately 9 times better performance is shown by Rayleigh fading channel in contrast to Rician.

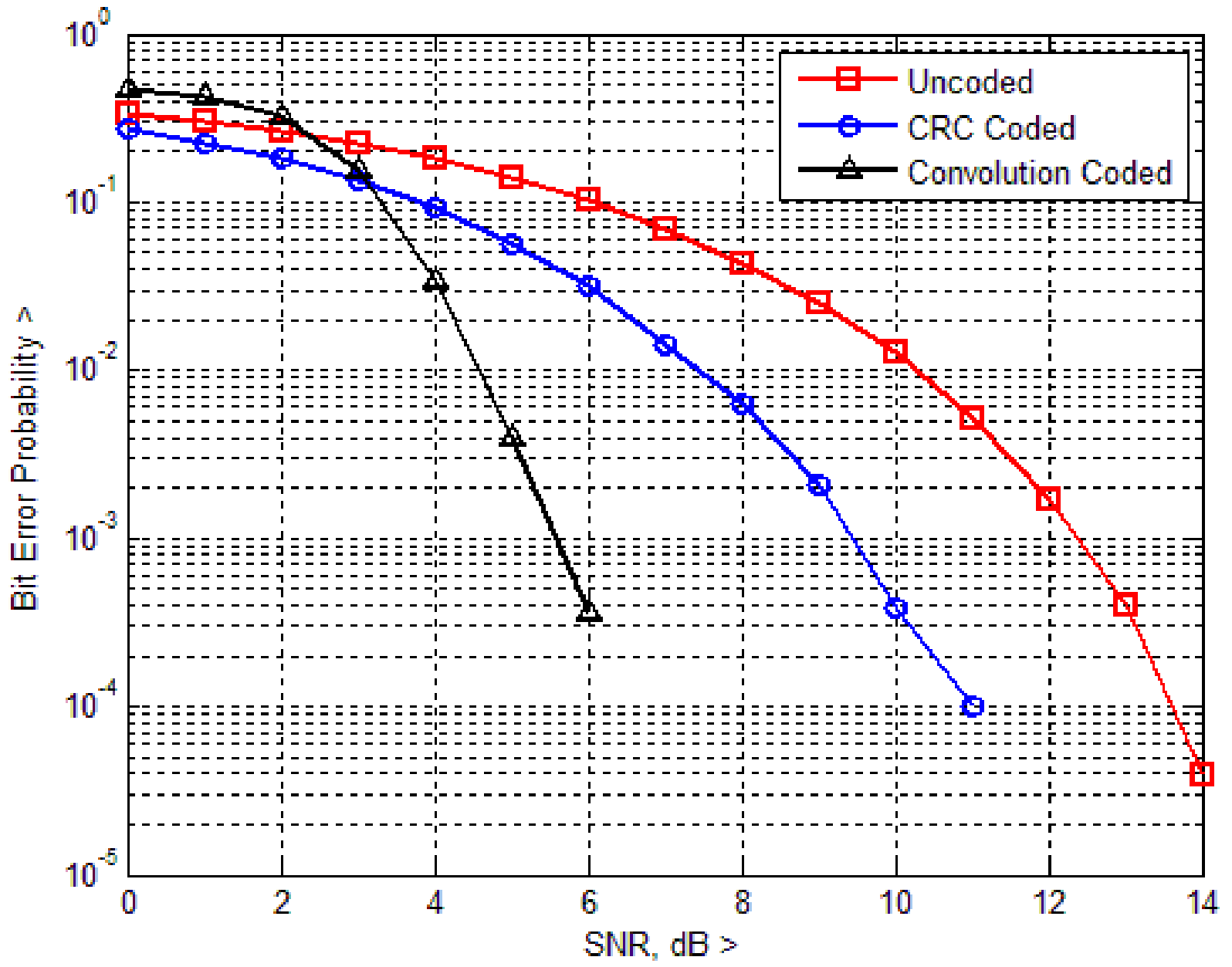

**Figure 5. Performance analysis between uncoded and interleaved coded OFDM system under AWGN channel.**

To improve the performance of this OFDM system, different interleaved FEC codes can be used. The most popular FEC codes are Cyclic Redundancy Check (CRC) and Convolutional code (CC) which are frequently used to achieve desire bit error probability and to transmit digital speech through wireless channels.

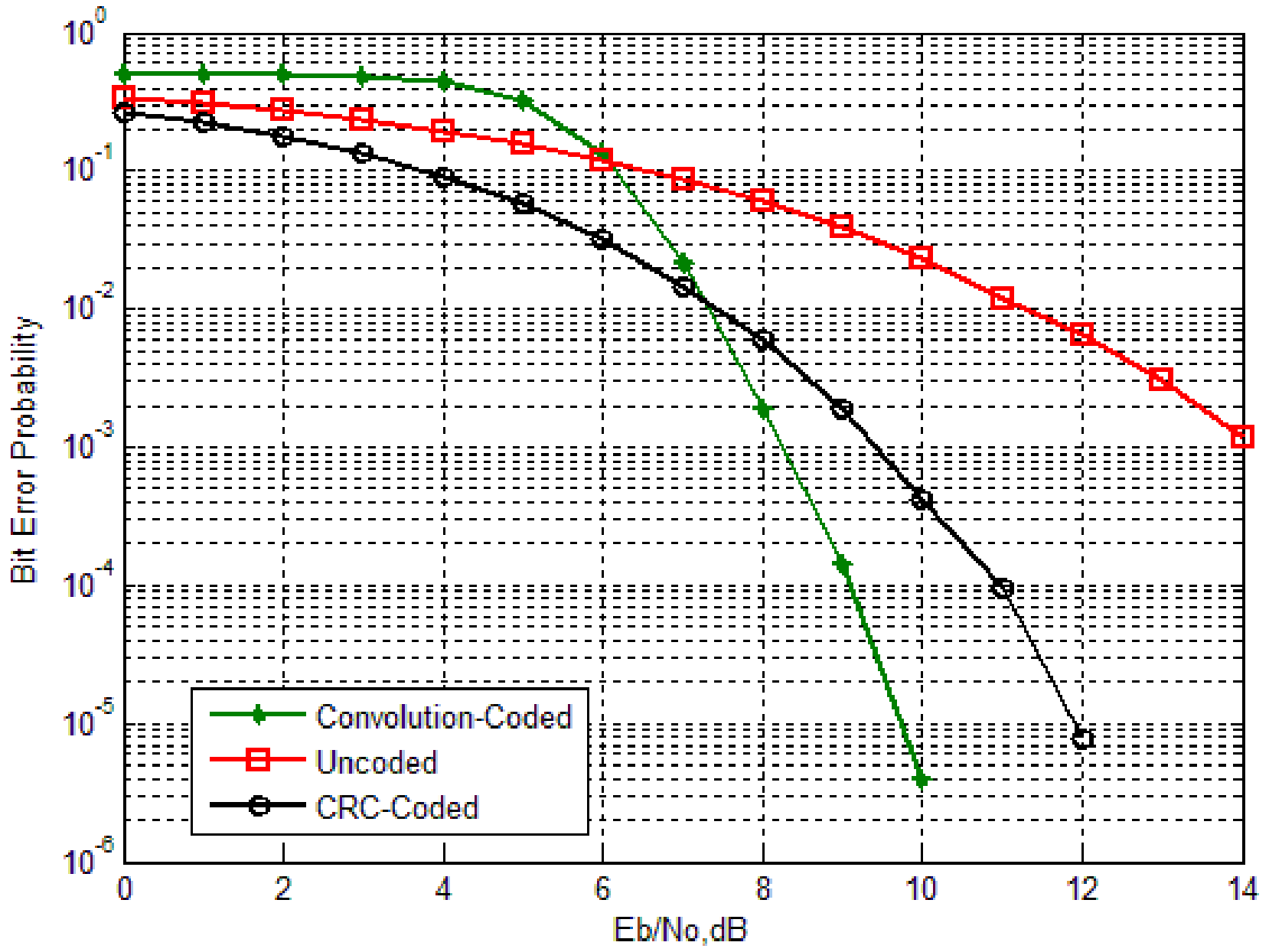

**Figure 6. Performance analysis between uncoded and interleaved coded OFDM system under Rayleigh channel.**

The performance comparison between coded and uncoded OFDM based wireless communication system under AWGN channel is presented by figure 5. It is seen from this figure that interleaved convolutional coding in OFDM can give performance improvement on AWGN channel over the uncoded as well as interleaved CRC coded OFDM system. We have measured the bit error probability of the OFDM system for SNR= 6 dB from our simulation are 0.0003594, 0.03136, and 0.1222 for using interleaved CC, interleaved CRC, and uncoded, respectively. It is remarkable here that till 3dB SNR, convolutional coded BER value does not strong than others but above 3dB SNR it showing better and better BER value. Contrast to CRC, CC shows approximately 2 order of magnitude better BER performance at a typical SNR value 6dB. Therefore, we can say that CC give better performance in AWGN channel.

Figure 6 demonstrates the improvement of our interleaved coded OFDM system under Rayleigh channel. It is clearly seen that desirable bit error probability is achieved when interleaved CC is used instead of interleaved CRC after SNR= 7 dB. It can be pointed that the bit error probability at SNR=9 dB is 0.0001406, 0.001875, 0.03851 for using interleaved CC coded, interleaced CRC coded, and uncoded OFDM system, respectively. Here, thirteen times better BER value is shown in CC coded system in contrast to CRC coded system. Therefore, we can conclude from the figure that the transmission of speech signal with considering Rayleigh fading effect, interleaved convolutional code provides better reproducibility of the output speech signal than interleaved CRC code. On the other hand, CRC coded system shows better BER value at the lower SNR region in all cases. The SNR value is below 3 and 7 for AWGN and fading channel, respectively where CRC system is outperform.

Finally, we investigate the impact of coding (CC and CRC) on OFDM system with Rician fading effect. Figure 7 reveals the effectiveness of convolutional coder in our considered OFDM system. A desirable bit error probability (0.0001836) is obtained when CC is used where the value of SNR is 9dB. This is the lowest bit error probability in same SNR among other coded system. So it is noticed here that interleaved CC based OFDM system explore better performance with Rician fading effect for speech signal transmission.

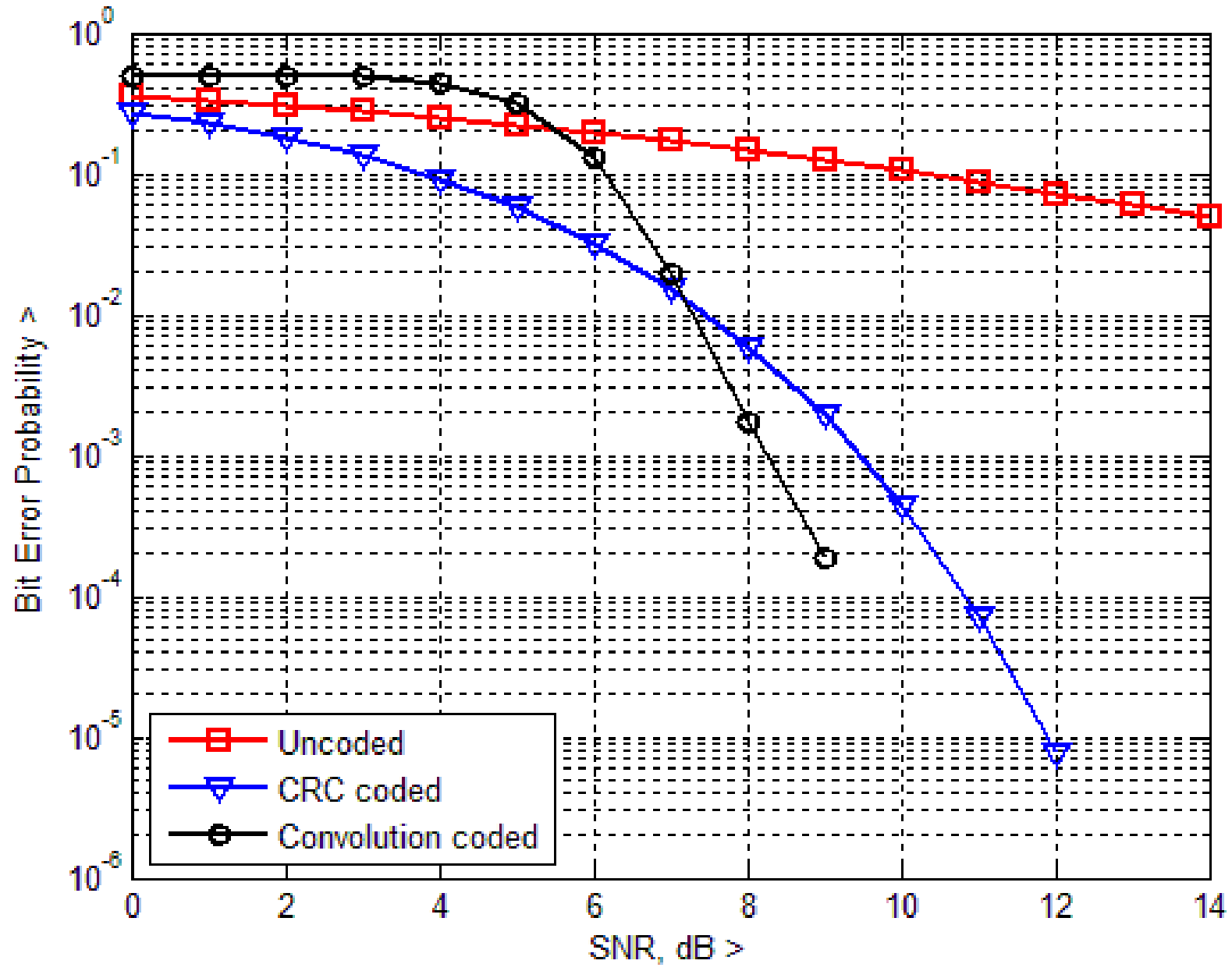

**Figure 7. Performance analysis between uncoded and interleaved coded OFDM system under Rician channel.**

In all simulation, it is observed that the convolutional code demonstrates better performance in contrast to others. This is because the process of code generation is different which reduce the delay of data transmission. In a CRC code, a block of n bits is generated by the encoder for an input block of k bits. On the other hand, an output block of n bits generated by the convolutional encoder depends not only on an input block of k bits but also on the block of (k-1) bits within a previous span of N-1 time unit. It also converts the entire data stream into one single codeword. Nevertheless, convolutional coding with a ½ rate and a constraint length of 7, a QPSK signal can be transmitted with less power. This can increases data rate for the same transmitter power and antenna size. However, it also increases of bandwidth by a factor of 2. In convolution code, generally the bandwidth expansion factor is defined by simply $n/k$, where $k/n$ is the code rate expressed as a ratio of the number of bits in the convolution encoder ($k$) to the number of channel symbols output by the convolution encoder ($n$) in a given encoder cycle.

## 6. Conclusion

The tremendous worldwide demand for high-speed mobile wireless communications is rapidly growing. OFDM technology promises to be a key technique for achieving the high data capacity and spectral efficiency requirements for wireless communication systems of the near future. As a result, the effect of FEC codes on considered OFDM system under AWGN as well as fading channels has been investigated. The performance can be improved by applying FEC codes in contrast to uncoded system. It is seen that interleaved FEC coded OFDM system combat against the noise in channels to achieve better performance. From simulation results it is observed that interleaved convolutional encoded OFDM system outperforms than that of interleaved cyclic redundancy check encoded OFDM system both AWGN channel as well as fading channels.

## References

[1] S. Hara and R. Prasad, “Multicarrier Techniques for 4G Mobile Communications”, Artech House, first edition, 2003.

[2] M. Engels, “Wireless OFDM Systems: How To Make Them Work?”, Kluwer Academic Publishers, 2002.

[3] V.Jagan Naveen, K.Murali Krishna, K.RajaRajeswari, “performance analysis of mc-cdma and OFDM in wireless rayleigh channel”, International Journal of Advanced Science and Technology, Vol. 5, December, 2010 .

[4] B. Lu, X. Wang, and K. R. Narayanan, “LDPC-based space-time coded OFDM systems over correlated fading channels: Performance analysis and receiver design," IEEE Trans. Wireless Commun., vol. 1, pp. 213-225, Apr. 2002.

[5] H. Kim, “Turbo coded orthogonal frequency division multiplexing for digital audio broadcasting", IEEE Intern. Confer. on Commun., vol. 1, pp. 420-424, 2000.

[6] C. Berrou and A. Glavieux, “Near optimum error correcting coding and decoding: Turbo-codes,” IEEE. Trans. Commun., Vol. 44, No. 10, pp. 1261 -1271, Oct. 1996.

[7] C. Athaudage and K. Sathananthan, “Probability of Error of SpaceTime Coded OFDM Systems with Frequency Offset in FrequencySelective Rayleigh Fading Channels,” in IEEE International Conference on Communications , vol. 4, pp. 2593– 2599, Seoul Korea, 16-20 May 2005.

[8] W. Hwang, H. Kang, and K. Kim, “Approximation of SNR Degradation Due to Carrier Frequency Offset for OFDM in Shadowed Multipath Channels,” IEEE Communications Letters , vol. 7, no. 12, pp. 581–583, December 2003.

[9] L. Rugini and P. Banelli, “BER of OFDM Systems Impaired by Carrier Frequency Offset in Multipath Fading Channels,” IEEE Transactions on Wireless Communications, vol. 4, no. 5, pp. 2279–2288, September 2005.

[10] C. Athaudage and K. Sathananthan, "Probability of Error of Space-Time Coded OFDM Systems with Frequency Offset in Frequency-Selective Rayleigh Fading Channels," in IEEE International Conference on Communications, vol. 4, Seoul Korea, 16-20 May 2005, pp. 2593–2599.

[11] Vivek K. Dwivedi, Abhinav Gupta, Richansh Kumar and G. Singh, "Performance Analysis of Co ded OFDM System Using Various Co ding Schemes" Progress In Electromagnetics Research Symposium Proceedings, Moscow, Russia, August 18-21, 2009.

[12] A. G. Burr and G. P. White, "Performance of Turbo-coded OFDM", The IEEE, Savoy Place, London WC2R 0BL, 1999. [13] Zahid Hasan,"Color Image Transmission: Error Analysis of a concatenated FEC Scheme Based OFDM system", In press Journal of Mobile communication, Vol:4, Issue:3, Page:81-85, ISSN:1990-794X, Pakistan. ,2010.

[14] M. K. Gupta, Vishwas Sharma," To improve bit error rate of turbo coded OFDM transmission over noisy channel" In press, Journal of Theoretical and Applied Information Technology, Vol:8, No:2, Page:162-168,Pakistan

[15] M. D. Haque, S. E. Ullah, M. M. Rahman, and M. Ahmed, "BER Performance Analysis of a Concatenated Low Density Parity Check Encoded OFDM System in AWGN and Fading Channels", Journal of Scientific Research, JSR Publication. Vol. 2, No. 1, PP. 46-53, 2010, Bangladesh, DOI: 10.3329/jsr.v2i1.2724.

[16] Md. Golam Rashed, Hasnat Kabir, Sk. Enayet Ullah and Rubaiyat Yasmin, "Performance evaluation of CRC Interleaved QPSK based Wireless Communication System for Color Image Transmission", Journal of Bangladesh Electronics Society (BES), Vol. 9, No. 1-2, PP. 151-159, Dhaka, Bangladesh, December 2009.

[17] Philip Koopman, Tridib Chakravarty," Cyclic Redundancy Code (CRC) Polynomial Selection For Embedded Networks" The International Conference on Dependable Systems and Networks, DSN-2004.

[18] Dennis Roddy, "Satellite Communcations" Third edition,McGraw-Hill Telecom Engineering. ISBN: 0-07-138285-2.DOI:10.1036/0071382852.

[19] B. Vucetic, J. Yuan (2000). Turbo codes: principles and applications. Springer Verlag. ISBN 978-0792378686.

[20] "Digital Video Broadcast (DVB); Frame structure, channel coding and modulation for a second generation digital terrestrial television broadcasting system (DVB-T2)". En 302 755 (ETSI) (V1.1.1). September 2009.